\documentclass[letterpaper, 10 pt, onecolumn]{ieeeconf}  

\IEEEoverridecommandlockouts                             
\usepackage{amsmath}  
\usepackage{amssymb}

\usepackage{graphicx}
\usepackage[space]{grffile}
\usepackage{cite}

\usepackage[top=1in, bottom=1in, left=1in, right=1in]{geometry}
\usepackage{wrapfig} 

\title{\LARGE \bf
Theoretical and Numerical Analysis of Approximate Dynamic Programming with Approximation Errors
}

\author{Ali Heydari$^1$
\thanks{$^{1}$Assistant Professor of Mechanical Engineering, South Dakota School of Mines and Technology, Rapid City, SD 57701, email: ali.heydari@sdsmt.edu. }}

\newtheorem{Thm}{Theorem} 
\newtheorem{Lem}{Lemma} 
\newtheorem{Rem}{Remark} 
\newtheorem{Def}{Definition} 
\newtheorem{Assumption}{Assumption} 

\DeclareMathOperator*{\argmin}{arg\,min}

\begin{document} 

\maketitle
\pagenumbering{arabic}
\pagestyle{plain}
\thispagestyle{plain}

\begin{abstract} 
This study is aimed at answering the famous question of how the approximation errors at each iteration of Approximate Dynamic Programming (ADP) affect the quality of the final results considering the fact that errors at each iteration affect the next iteration. To this goal, convergence of Value Iteration scheme of ADP for deterministic nonlinear optimal control problems with undiscounted cost functions is investigated while considering the errors existing in approximating respective functions. The boundedness of the results around the optimal solution is obtained based on quantities which are \textit{known} in a general optimal control problem and assumptions which are \textit{verifiable}. Moreover, since the presence of the approximation errors leads to the deviation of the results from optimality, sufficient conditions for stability of the system operated by the result obtained after a \textit{finite} number of value iterations, along with an estimation of its \textit{region of attraction}, are derived in terms of a \textit{calculable} upper bound of the control approximation error. Finally, the process of implementation of the method on an orbital maneuver problem is investigated through which the assumptions made in the theoretical developments are verified and the sufficient conditions are applied for guaranteeing stability and near optimality.
\end{abstract}

\section{Introduction}

Approximate (or adaptive) dynamic programming (ADP) or reinforcement learning (RL) has been investigated extensively by different researchers as a powerful tool for approximating solutions to mathematically intractable problems seeking optimum, \cite{Watkins, Werbos, Sutton, Bertsekas_NDP, Bala_Biega, Prokhorov, Venaya, Si, AlTamimi, Zhao_Xu_Jagannathan, Liu_TCYB, Heydari_NN_FinHor}.
ADP has shown its great potential in aerospace applications as well, \cite{Bala_Biega, Bala_Han_JGCD, Ferrari_Stengel, JonathanHow_ADP, Bala_Jie, Heydari_JGCD_Rendezvous}, from control of agile missiles to spacecraft rendezvous.
The most popular algorithm for ADP is \textit{Value Iteration} (\textit{VI}), \cite{Sutton, LewisContSystMag}. 
The convergence of VI for problems subject to this study, i.e., problems with undiscounted cost functions and continuous state and action spaces, was analyzed in \cite{Landelius_PhDThesis, Bala_Liu_Convergence} for linear systems and in \cite{Lincol_RelaxingDynProg}, \cite{AlTamimi}, \cite{Heydari_TCYB} for nonlinear systems. A crucial assumption in the cited convergence proofs is \emph{perfect} function approximation, i.e., neglecting function approximation errors. However, this assumption rarely holds in nonlinear problems. The concern with the existence of the approximation errors is due to the fact that the errors \emph{propagate} throughout the iterations, i.e., their consequences may grow in future iterations. In other words, a `resonance' type phenomenon may happen, regardless of how small each single error term is, which would lead to unreliability of the solution. 

Incorporating the approximation errors in the analysis of VI, i.e., analyzing \textit{approximate VI} (\textit{AVI}), is a challenging area and the appeared results, to the best of the knowledge of the author, are limited to \cite{Bertsekas_NDP, Singh_Discounted_ADP_ErrorAnalysis, Szepesv_AVI_API_ErrorAnalysis, Szepesv_AVI_ErrorAnalysis, Liu_TCYB}. Problems with \textit{discounted} cost functions were the subject of Refs. \cite{Bertsekas_NDP, Singh_Discounted_ADP_ErrorAnalysis, Szepesv_AVI_API_ErrorAnalysis, Szepesv_AVI_ErrorAnalysis}. These results however, are not extendable to undiscounted cost-functions, because, the `forgetting' nature of discounted problems plays a critical role in the derivations and if the discount factor converges to one, i.e., the problem becomes undiscounted, the developed bounds go to infinity.
Ref. \cite{Liu_TCYB}, however, investigated AVI for undiscounted cost functions and provided some interesting results. However, the utilized assumptions are relatively more restrictive and not easily verifiable, compared to this study. Assuming the approximation errors can be written in a multiplicative form, instead of an additive form, i.e., assuming $\hat{V}(x) \leq \sigma V(x)$ holds uniformly for some positive constant $\sigma$, instead of assuming $\hat{V}(x) = V(x) + \epsilon(x)$, for a function $\epsilon(.)$, where $V(.)$ and $\hat{V}(.)$ denote the exact and the approximated functions, is one of such assumptions. Moreover, the developed results require $\sigma$ to be upper bounded by a term which involves the \textit{optimal} value function. As for non-value iteration based approaches in which the approximation errors are not neglected, interested readers are referred to \cite{Vamvoudakis_Automatica, Dierks_Jagannathan_Approx_Error, Vamvoudakis_Vrabie_Lewis}.

Considering the scarcity of the available studies, the prevalence of approximation errors, and the dramatic success of value iteration in solving optimal control problems in different applications, including aerospace systems \cite{Bala_Biega, Ferrari_Stengel, JonathanHow_ADP, Bala_Jie}, rigorous theoretical analyses in the area are of interest to the controls community. 
This study is aimed at this pursuit, i.e., contributing to the mathematical rigor of the field of intelligent control, more specifically, ADP for control. 
This is done through developing a sufficient condition for boundedness/convergence of the iterations under the presence of the approximation errors. The sufficient condition can be easily checked for general deterministic nonlinear systems. Moreover, the stability of the system operated using the approximate solution, obtained through a \textit{finite} number of iterations of AVI, is investigated and required conditions for guaranteeing stability are derived in terms of known and calculable parameters for general systems. It should be noted that the presence of the approximation errors and also possibly immature conclusion of iterations not only lead to the deviation of the results from optimality, but also, can potentially lead to instability/unreliability of the system operated using the resulting solution, which may lead to catastrophic outcomes when utilized in aerospace systems. Therefore, investigation of the stability and deriving sufficient conditions for guaranteeing stability are required. Moreover, the important concern that a neurocontroller is valid only when the state trajectory remains within the domain for which the controller is trained is addressed through finding an estimation of the \textit{region of attraction} (ROA) for the result obtained through the AVI. 
It should be noted that in the general case, if a neurocontroller is trained for a given domain, it is \textit{not} guaranteed that any state trajectory initiated from the domain remains inside the domain. If it exists the domain, then the neurocontroller becomes invalid. However, once an estimation of the ROA is found any state trajectory initiated from the domain, will remain within the domain and therefore, the controller remains valid for use.
Finally, interested readers are referred to \cite{Heydari_SAVI} for some recent developments of the author regarding \textit{stablizing value iteration}, i.e., the exact or approximate VI which is initiated from a stabilizing initial guess. Such a scheme guarantees the stability of the system during \textit{online learning}.
\footnote{It must be added that the current version of this study has overlaps with the first version of \cite{Heydari_SAVI} on Theorem \ref{Thm_Cont_Vi} and Lemma \ref{Lem_Boundedness}.}

The rest of this study is organized as follows. The optimal control problem is presented in Section II and the exact VI scheme is revisited in Section III. Section IV presents the approximate VI, followed by the theoretical analyses in section V. Afterward a famous aerospace example is numerically investigated in Section VI. Finally, concluding remarks are given in Section VII.

\section{Optimal Control Problem} \label{ProblemFormulation}
The nonlinear discrete-time dynamics of the system are assumed to be given by 
\begin{equation}
x_{k+1}=f(x_k,u_k), k \in \mathbb{N}, \label{Dynamics}
\end{equation}
where $x$ and $u$ are the state and control vectors, respectively, with the dimensions of $n$ and $m$. Function $f:\mathbb{R}^n \times \mathbb{R}^m \to \mathbb{R}^n$ is assumed to be smooth versus its inputs and $f(0,0)=0$. Sub-index $k$ is the time-index and $\mathbb{N}$ denotes the non-negative integers. The problem is defined as finding a \textit{feedback control policy} $h:\mathbb{R}^n \to \mathbb{R}^m$, i.e., $u_k = h(x_k)$, such that the cost function given below is minimized subject to dynamics (\ref{Dynamics}) and given any initial conditions $x_0$.
\begin{equation}
J=\sum_{k=0}^\infty {U(x_k,u_k)}. \label{CostFunction}
\end{equation}
It is assumed that $U(x_k,u_k):=Q(x_k) + u_k^TRu_k$ for a convex and smooth positive definite function $Q:\mathbb{R}^n \to \mathbb{R}_+$ and a positive definite $m \times m$ real matrix $R$. The set of non-negative reals is denoted with $\mathbb{R}_+$. 

Let the \emph{cost-to-go} or \emph{value function} of a control policy $h(.)$, denoted by $V_h:\mathbb{R}^n \to \mathbb{R}_+$, be defined as 
\begin{equation}
V_h(x_0)=\sum_{{k}=0}^\infty {U\big(x_{k}^h, h(x_{k}^h)\big)}. \label{ValueFunction_of_h}
\end{equation}
In (\ref{ValueFunction_of_h}) one has $x_k^h:=f\big(x_{k-1}^h,h(x_{k-1}^h)\big), \forall k \in \mathbb{N}-\{0\},$ and $x_0^h := x_0$, i.e., $x_k^h$ denotes the $k$th element on the state history initiated from $x_0$ and generated using $h(.)$. 

\begin{Def} \label{Def1}
An \emph{admissible} control policy within a set is defined as a control policy which is a continuous function and leads to an upper bounded value function for any $x_0$ in the set. 
\end{Def}

\begin{Rem} 
The defined admissibility is different from the usual definition, including \cite{AlTamimi}, in the sense that the control policy is not required to asymptotically stabilize the system \cite{Khalil} or to have $h(0)=0$ besides being continuous and leading to a finite value function. However, the assumed two features (continuity and finiteness of the value function) lead to those (not explicitly assumed) characteristics, as the tail of a convergent series can be made arbitrarily small (cf. p. 59 \cite{Rudin}).
\end{Rem} 

\begin{Assumption} \label{Assum_ExistingAdmissibleCont}
Considering $\Omega \subset \mathbb{R}^n$ as a compact set containing the origin, there exists at least one admissible control policy for the system within $\Omega$.
\end{Assumption}

This assumption is made for guaranteeing that there is no state vector in $\Omega$ for which the value function associated with the \textit{optimal} control policy is infinite. Because, otherwise, the optimal control policy will not be `optimal' compared with the existing admissible control policy.

\section{Exact Value Iteration}
The value function of a policy $h(.)$ satisfies
\begin{equation}
V_h(x)= U\big(x,h(x)\big) + V_h\Big(f\big(x,h(x)\big)\Big), \forall x \in \mathbb{R}^n, \label{Recursive_CostToGo}
\end{equation}
based on Eq. (\ref{ValueFunction_of_h}). Let the \textit{optimal value function}, denoted with $V^*(.)$, be defined as the value function of the optimal control policy. The optimal value function satisfies the Bellman equation \cite{Kirk}
\begin{equation}
		h^*(x) = \argmin_{u\in\mathbb{R}^m} \Big( U\big(x,u\big) + V^*\big(f(x,u)\big)\Big), \label{Bellman_eq2}
\end{equation}
\begin{equation}
		V^*(x) = \min_{u\in\mathbb{R}^m} \Big( U(x,u) + V^*\big(f(x,u)\big)\Big). \label{Bellman_eq1}
\end{equation}
However, the famous \textit{curse of dimensionality} \cite{Kirk} leads to the intractability of the approach of using Bellman Eq. for solving the problem in the general case. 
The idea in ADP is \textit{approximating} the optimal value function for remedying the problem of curse of dimensionality. The approximation is typically done using look-up tables or function approximators, e.g., neural networks. \textit{Critic}, in the ADP/RL literature, is the term used for the optimal value function approximator.
One selects a set, called the \textit{domain of interest}, which is compact, connected, and contains the origin, within which the value function will be approximated.
Denoting the domain of interest with $\Omega$, it needs to be selected based on the given system and its operation envelope, as the ADP solution is valid only if the state trajectory entirely remains within $\Omega$. 

Value iteration (VI) is one of the learning schemes for finding the optimal value function. 
The VI process starts with an initial guess $V^0(.)$ and iterates through 
\begin{equation}
		V^{i+1}(x) = \min_{u} \Big( U(x,u) + V^i\big(f(x,u)\big)\Big), \forall x \in \Omega, \label{VI_ValueUpdate}
\end{equation}
for $i=0,1,...$ until the iterations converge. As one of the available convergence proofs, \cite{Heydari_TCYB} shows that if the initial guess on $V^0(.)$ is smooth and $0 \leq V^0(x) \leq U(x,0), \forall x \in \Omega$, then the VI converges monotonically to the optimal solution. Utilizing the converged value function, denoted with $V^*(.)$, the optimal control policy can be obtained using Eq. (\ref{Bellman_eq2}).

\section{Approximate Value Iteration} \label{AVI}
VI is based on the assumption that one can \emph{exactly} approximate/reconstruct the right hand side of Eq. (\ref{VI_ValueUpdate}). However, this is rarely the case for general nonlinear problems. 
Parametric function approximators are used in practice for this purpose. The approximation process leads to some approximation errors. Incorporating the errors, Eq. (\ref{VI_ValueUpdate}) leads to
\begin{equation}
		\hat{V}^{i+1}(x) = \min_{u} \Big( U(x,u) + \hat{V}^i\big(f(x,u)\big)\Big) + \epsilon^i(x), \forall x \in \Omega. \label{AVI_1}
\end{equation}
Function $\epsilon^i(.)$, in (\ref{AVI_1}), denotes the approximation error at the $i$th iteration and $\hat{V}^i(.)$ denotes the \textit{approximate value function} at this iteration.
It should be noted that the right hand side of Eq. (\ref{AVI_1}) also contains an approximate quantity, generated from the previous iteration. 

A critical point is the fact that $\epsilon^i(x)$ does \emph{not} represent the error between the \emph{exact} and the \emph{approximate} value functions, denoted with $V^{i+1}(x)$ and $\hat{V}^{i+1}(x)$, respectively. The exact value function $V^{i+1}(x)$ is based on using the exact $V^{i}(x)$ in the right hand side of (\ref{VI_ValueUpdate}), while, $\hat{V}^{i+1}(x)$ is being calculated based on $\hat{V}^{i}(x)$, per (\ref{AVI_1}). The difference between $V^{i+1}(x)$ and $\hat{V}^{i+1}(x)$ is an approximation error which is the cumulative effect of $\epsilon^i(.)$'s in the previous iterations. The `per iteration' error, denoted with $\epsilon^i(x)$, however, is simply the error of approximating/replacing $\min_{u} \Big( U(x,u) + \hat{V}^i\big(f(x,u)\big)\Big)$ with $\hat{V}^{i+1}(x)$.
Also, note that when $\epsilon^i(.) \neq 0$, the convergence of the approximate VI (AVI) does not follow from the cited previous investigations, as mentioned in the introduction. 

It should be mentioned that one typically trains a control approximator (actor) as well at each iteration of AVI, to approximate the solution to the minimization problem given by (\ref{Bellman_eq2}), in which $V^*(.)$ is replaced with $\hat{V}^i(.)$. The actor will give rise to \textit{another} approximation error term in the solution process, as seen in \cite{Liu_TCYB}. 
However, the effect of the actor's approximation error can be removed from the convergence analysis of AVI, as the actor training can be postponed till after the conclusion of the value function learning through Eq. (\ref{VI_ValueUpdate}) or (\ref{AVI_1}) in \textit{offline} learning. In other words, one can learn the optimal value function and then use the result for training the actor. The detailed algorithm is presented in \cite{Heydari_TCYB}. However, one might be interested in \textit{online} learning, as it leads to the feature of not needing the perfect knowledge of the internal dynamics of the system \cite{AlTamimi, Werbos2012}. Even in case of online learning the effect of the actor's approximation error can be removed from the convergence analysis, as the control will be directly calculated from the minimization of the right hand side of Eq. (\ref{AVI_1}) and applied on the system. 
The point is, the actor's approximation accuracy does not affect the critic training, even though the actor will be updated simultaneously along with the critic in online learning. Of course, once the offline or online learning is concluded, the system will be operated using the control resulting from the trained actor, hence, the stability of the system could be at risk due to the actor's approximation errors. After the convergence analysis, this concern will be investigated in this study.

\section{Theoretical Analyses}

\subsection{Continuity Analysis}
Smooth function approximators are shown to uniformly approximate a function if the function is \textit{continuous}, \cite{Weierstrass_Theorem,Hornik_NN_Continuity}. Otherwise, the approximation accuracy is not guaranteed to be suitable on \textit{new} states which were \textit{not} used in the training. On the other hand, the minimization operation in (\ref{AVI_1}) may lead to discontinuity of the right hand side versus $x$, since, the $u$ which minimizes the term is given by
\begin{equation}
		u = \argmin_{u} \Big( U(x,u) + \hat{V}^{i}\big(f(x,u)\big)\Big), \label{AVI_Control}
\end{equation}
and hence, may change discontinuously, since $\argmin(.)$ is \textit{not} a continuous function generally. Therefore, an important step is analyzing the continuity of the function subject to approximation, that is,
\begin{equation}
		\mathcal{V}^{i+1}(x) := \min_{u} \Big( U(x,u) + \hat{V}^i\big(f(x,u)\big)\Big), \forall x \in \Omega. \label{AVI_1_approx}
\end{equation}
Note that the the difference between $\mathcal{V}^{i+1}(.)$ and $\hat{V}^{i+1}(.)$ is the fact that the latter is the approximation of the former, i.e., $\hat{V}^{i+1}(.)= \mathcal{V}^{i+1}(.) + \epsilon^i(.)$. 

Let $\mathcal{C}({x})$ (respectively, $\mathcal{C}(\Omega)$) denote the set of continuous functions at point ${x}$ (respectively, within $\Omega$). The following theorem establishes the desired continuity.

\begin{Thm} \label{Thm_Cont_Vi} If the approximate value iteration scheme, implemented using a continuous function approximator, is initiated using a continuous initial guess, then the function subject to approximation by the critic will be continuous at any finite iteration.
\end{Thm}
\textit{Proof}: 
Based on the continuity of the function approximator, one has $\hat{V}^i(.) \in \mathcal{C}(\Omega), \forall i$. The theorem can be proved by showing that if $\hat{V}^i(.) \in \mathcal{C}(\Omega)$ then $\mathcal{V}^{i+1}(.) \in \mathcal{C}(\Omega)$.
Let $W(x,u) := U(x,u) + \hat{V}^i\big(f(x,u)\big)$ and $h(x) = \argmin_{u\in\mathbb{R}^m} W(x,u)$. Note that functions $f(.,.)$ and $U(.,.)$ are smooth, hence, continuous. Since, $W(.,h(.))=\mathcal{V}^{i+1}(.)$ the proof of continuity of $W\big(.,h(.)\big)$ suffices. The proof is done by showing that the directional limit of $W\big(.,h(.)\big)$ at any selected point is equal to its evaluation at the point, and hence, it is continuous at that point (motivated by \cite{Heydari_Franklin}).

Let $\bar{x}$ be an arbitrary point in $\Omega$. Set
\begin{equation}
\bar{u}:=h(\bar{x}). \label{Cont_Thm_eq1}
\end{equation}
Select an open set $\alpha \subset \mathbb{R}^n$ such that $\bar{x}$ belongs to the boundary of $\alpha$ and limit
\begin{equation}
\hat{u} := \lim_{x \to \bar{x}, x \in \alpha} h(x), \label{Cont_Thm_eq2}
\end{equation}
exists. If $\bar{u} = \hat{u}$, for every such $\alpha$, then $h(.) \in \mathcal{C}(\bar{x})$. In this case the continuity of $W\big(.,h(.)\big)$ at $\bar{x}$ follows from the continuity of its forming functions, \cite{Rudin}. 

Now assume $\bar{u} \neq \hat{u}$, for some $\alpha$ denoted with $\alpha_0$. From $W(.,\hat{u}) \in \mathcal{C}(\Omega)$ for the given $\hat{u}$, one has
\begin{equation}
W(\bar{x},\hat{u}) = \lim_{x \to \bar{x}, x \in \alpha_0} W(x,\hat{u}), \label{Cont_Thm_eq4}
\end{equation}
If it can be shown that, for every selected $\alpha_0$, one has
\begin{equation}
W(\bar{x},\bar{u}) = W(\bar{x},\hat{u}), \label{Cont_Thm_eq5}
\end{equation}
then the continuity of $W\big(.,h(.)\big)$ at $\bar{x}$ follows, because from (\ref{Cont_Thm_eq4}) and (\ref{Cont_Thm_eq5}) one has
\begin{equation}
W(\bar{x},\bar{u}) = \lim_{x \to \bar{x}} W(x,\hat{u}),
\label{Cont_Thm_eq6}
\end{equation}
and (\ref{Cont_Thm_eq6}) leads to the continuity by definition, \cite{Rudin}. 

\begin{wrapfigure}{r}{0.2\textwidth}
  \vspace{-20pt}
  \begin{center}
		\includegraphics[width=0.15\columnwidth]{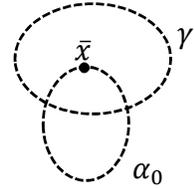}
  \end{center}
  \vspace{-10pt}
		\caption{Schematic of point $\bar{x}$ and open sets $\alpha_0$.} 
		\label{Fig_alpha_gamma_sets}
  \vspace{-10pt}
\end{wrapfigure}

The proof that (\ref{Cont_Thm_eq5}) holds is done by contradiction. 
Assume that for some $\bar{x}$ and some $\alpha_0$ one has
\begin{equation}
W(\bar{x},\bar{u}) > W(\bar{x},\hat{u}). \label{Cont_Thm_eq9}
\end{equation}
Inequality (\ref{Cont_Thm_eq9}) leads to $h(\bar{x}) \neq \bar{u}$. But, this is against (\ref{Cont_Thm_eq1}), hence, (\ref{Cont_Thm_eq9}) cannot hold.
Now, assume 
\begin{equation}
W(\bar{x},\bar{u}) < W(\bar{x},\hat{u}), \label{Cont_Thm_eq7}
\end{equation}
hence there exists some $\epsilon_1 > 0$ such that
\begin{equation}
W(\bar{x},\bar{u}) + \epsilon_1 = W(\bar{x},\hat{u}), \label{Cont_Thm_eq7_1}
\end{equation}
then, due to the continuity of both sides of (\ref{Cont_Thm_eq7_1}) at $\bar{x}$ for the fixed $\bar{u}$ and $\hat{u}$, there exists an open set $\gamma$ containing $\bar{x}$, see Fig. \ref{Fig_alpha_gamma_sets}, and some $\epsilon_2>0$, such that 
\begin{equation}
W(x,\bar{u})  + \epsilon_2 < W(x,\hat{u}),\forall x \in \gamma. \label{Cont_Thm_eq8}
\end{equation}

Given $W\big(x,h(x)\big) \leq  W(x,\bar{u})$, inequality (\ref{Cont_Thm_eq8}) implies that at points which are \textit{close enough} to $\bar{x}$, function $W\big(x,h(x)\big)$ is away from $W(x,\hat{u})$ at least by a margin of $\epsilon_2$. But, this contradicts Eq. (\ref{Cont_Thm_eq2}) which, 
implies that $h(x)$ can be made arbitrarily close to $\hat{u}$ as $x$ gets close to $\bar{x}$ within $\alpha_0$. The reason is, the latter, given the continuity of $W(x,u)$ versus both $x$ and $u$, leads to the conclusion that function $W\big(x,h(x)\big)$ can be made arbitrarily close to $W(x,\hat{u})$ if $x$ approaches $\bar{x}$ from a certain direction. Note that sets $\gamma$ and $\alpha_0$ are not disjoint, as $\bar{x}$ is \emph{within} $\gamma$ and on the \emph{boundary} of $\alpha_0$, as shown in Fig. \ref{Fig_alpha_gamma_sets}. Hence, inequality (\ref{Cont_Thm_eq7}) also cannot hold. 
Therefore, (\ref{Cont_Thm_eq5}) holds and hence, $W(.,h(.)) \in \mathcal{C}(\bar{x})$. Finally, the continuity of the function subject to investigation at any arbitrary $\bar{x} \in \Omega$, leads to the continuity of the function in $\Omega$.
\QED

\subsection{Convergence Analysis}
Analysis of boundedness and convergence of sequence $\{ \hat{V}^i(x) \}_{i=0}^\infty$ resulting from the approximate VI given by Eq. (\ref{AVI_1}) and its relation versus the optimal value function is presented in this subsection.     
Define $\{ \overline{V}^i(x) \}_{i=0}^\infty$ and $\{ \underline{V}^i(x) \}_{i=0}^\infty$ where $\overline{V}^i:\mathbb{R}^n\to\mathbb{R_+}$ and $\underline{V}^i:\mathbb{R}^n\to\mathbb{R_+}$ as sequences of functions initiated from some $\overline{V}^0(.)$ and $\underline{V}^0(.)$ and propagated using  
\begin{equation}
		\overline{V}^{i+1}(x) = \min_{u} \Big( U(x,u) +cU(x,0)+ \overline{V}^i\big(f(x,u)\big)\Big), \forall x \in \Omega, \label{V_Upper_1}
\end{equation}
\begin{equation}
		\underline{V}^{i+1}(x) = \min_{u} \Big( U(x,u) -cU(x,0)+ \underline{V}^i\big(f(x,u)\big)\Big), \forall x \in \Omega. \label{V_Lower_1}
\end{equation}
Now, assuming an upper bound for the approximation error $\epsilon^i(x)$
the following results can be obtained.
\begin{Lem} \label{Lem_Boundedness} Let $| \epsilon^i(x)| \leq c U(x,0), \forall i \in \mathbb{N}$ for some $c\in [0,1)$. If the recursive relations given by Eqs. (\ref{AVI_1}), (\ref{V_Upper_1}), and (\ref{V_Lower_1}) are initialized such that $\underline{V}^0(x) \leq \hat{V}^0(x) \leq \overline{V}^0(x), \forall x \in \Omega$, then, one has $\underline{V}^i(x) \leq \hat{V}^i(x) \leq \overline{V}^i(x), \forall x \in \Omega, \forall i\in \mathbb{N}$. Moreover, $\underline{V}^i(x)$ and $\overline{V}^i(x)$ are, respectively, the \textit{greatest} lower bound and the \textit{least} upper bound of $\hat{V}^i(x)$ if $\underline{V}^0(x)=\overline{V}^0(x)=\hat{V}^0(x)$.   
\end{Lem}

\textit{Proof}:
The lemma can be proved using mathematical induction. Initially $\underline{V}^0(x) \leq \hat{V}^0(x) \leq \overline{V}^0(x), \forall x \in \Omega$ by assumption. Let $\underline{V}^i(x) \leq \hat{V}^i(x) \leq \overline{V}^i(x), \forall x \in \Omega$ hold for some $i$. Comparing Eq. (\ref{V_Upper_1}) with Eq. (\ref{AVI_1}) it follows that $\hat{V}^{i+1}(x) \leq \overline{V}^{i+1}(x)$, since $ \epsilon^i(x) \leq c U(x,0)$ and $\hat{V}^i(x) \leq \overline{V}^i(x)$. Therefore, one has $\hat{V}^i(x) \leq \overline{V}^i(x), \forall i\in \mathbb{N}$. The proof of $\underline{V}^i(x) \leq \hat{V}^i(x), \forall i\in \mathbb{N}$ is similar through comparing Eq. (\ref{V_Lower_1}) with Eq. (\ref{AVI_1}) and using mathematical induction. Proof of the last part of the lemma follows from assuming $\epsilon^i(x) = c U(x,0), \forall i$ (respectively, $\epsilon^i(x) = -c U(x,0), \forall i$) which leads to $\hat{V}^i(x) = \overline{V}^i(x)$ (respectively, $\hat{V}^i(x) = \underline{V}^i(x)$). Therefore, there are no other `tighter' bounds for $\hat{V}^i(x)$.
$\QED$

It can be seen that functions $\overline{V}^i(.)$ and $\underline{V}^i(.)$ are the value functions at the $i$th iteration of \textit{exact VI} for cost functions 
\begin{equation}
\overline{J}=\sum_{k=0}^\infty \Big( U(x_k,u_k) + cU(x_k,0) \Big), \label{Cost_Upper_1}
\end{equation}
\begin{equation}
\underline{J}=\sum_{k=0}^\infty \Big( U(x_k,u_k) - cU(x_k,0) \Big), \label{Cost_Lower_1}
\end{equation}
respectively, subject to dynamics (\ref{Dynamics}), considering recursive relations (\ref{V_Upper_1}) and (\ref{V_Lower_1}). The following lemma provides the sufficient conditions for their convergence to the respective optimal value functions.
\begin{Lem} \label{Lem_ExactVI_Convergence} The exact value iterations given by Eqs. (\ref{V_Upper_1}) and (\ref{V_Lower_1}) converge to the optimal value functions of cost functions (\ref{Cost_Upper_1}) and (\ref{Cost_Lower_1}), respectively, if they are initialized by smooth functions $\underline{V}^0(.)$ and $\overline{V}^0(.)$ such that $0 \leq \underline{V}^0(x) \leq (1-c)U(x,0), \forall x \in \Omega$ and $0 \leq \overline{V}^0(x) \leq (1+c)U(x,0), \forall x \in \Omega$, where $c\in [0,1)$.
\end{Lem}

\textit{Proof}: The proof follows from \cite{Heydari_TCYB}, since, iterations given by (\ref{V_Upper_1}) and (\ref{V_Lower_1}) are exact VIs. 
$\QED$

Considering Lemmas \ref{Lem_Boundedness} and \ref{Lem_ExactVI_Convergence} the following theorem proves the boundedness of the elements of  $\{ \hat{V}^i(x) \}_{i=0}^\infty$ resulting from the approximate VI.

\begin{Thm} \label{Thm_Boundedness} Let $| \epsilon^i(x)| \leq c U(x,0), \forall x \in \Omega, \forall i \in \mathbb{N}$ for some $c\in [0,1)$. 
If the approximate value iteration given by Eq. (\ref{AVI_1}) is initialized such that $0 \leq \hat{V}^0(x) \leq (1-c)U(x,0), \forall x \in \Omega$, then, the elements of sequence $\{ \hat{V}^i(x) \}_{i=0}^\infty$ as $i \to \infty$ are \textit{bounded} by the optimal value functions of cost functions (\ref{Cost_Upper_1}) and (\ref{Cost_Lower_1}) denoted with $\overline{V}^*(x)$ and $\underline{V}^*(x)$, respectively, in the sense that the greatest lower bound of $\hat{V}^i(x)$ converges to $\underline{V}^*(x)$ and the least upper bound of $\hat{V}^i(x)$ converges to $\overline{V}^*(x)$ as $i \to \infty$.
  
\end{Thm}

\textit{Proof}:
The proof follows from the boundedness of $\{ \hat{V}^i(x) \}_{i=0}^\infty$ given in Lemma \ref{Lem_Boundedness} and the convergence of the bounds for smooth $\underline{V}^0(x)$ and $\overline{V}^0(x)$ which satisfy 
$0 \leq \underline{V}^0(x) = \hat{V}^0(x)= \overline{V}^0(x) \leq (1-c)U(x,0), \forall x \in \Omega$
based on Lemma \ref{Lem_ExactVI_Convergence}. 
$\QED$

Moreover, the following result can be achieved, with the uniformness feature which will be used in stability analysis.
\begin{Thm} \label{Thm_UniformConv_vs_c} Let $| \epsilon^i(x)| \leq c U(x,0), \forall x \in \Omega, \forall i \in \mathbb{N}$ for some $c\in [0,1)$. 
Also, let the approximate value iteration given by Eq. (\ref{AVI_1}) be initialized such that $0 \leq \hat{V}^0(x) \leq (1-c)U(x,0), \forall x \in \Omega$. 
As $c \to 0$, 
the results from the approximate value iteration (\ref{AVI_1}) converges \textit{uniformly} to the results from the exact value iterations given by (\ref{VI_ValueUpdate}) corresponding to cost function (\ref{CostFunction}) in compact set $\Omega$. More specifically, the least upper bound and the greatest lower bound of $\hat{V}^i(x)$ for $i \to \infty$ converge uniformly to the optimal value function associated with cost function (\ref{CostFunction}) as $c \to 0$. 
\end{Thm}
\textit{Proof}: The proof is given in the appendix.

Theorem \ref{Thm_Boundedness} proves that sequence $\{ \hat{V}^i(x) \}_0^\infty$ is upper and lower bounded. Then, Theorem \ref{Thm_UniformConv_vs_c} proves the uniform convergence of these bounds to the desired optimal solution if $c \to 0$. However, when the approximation error does not vanish, the mere fact that the sequence is upper bounded does not prove its convergence (the elements of a sequence can be upper bounded but oscillatory).
The established boundedness resembles the `convergence to a neighborhood' or interval presented in \cite{Liu_TCYB}, however, besides the idea behind the analysis which is different in here, the assumptions are also different and less restrictive in this study.

\subsection{Stability Analysis}
Even though it is proved that the AVI result remains bounded (Theorem \ref{Thm_Boundedness}), it is not necessarily optimal, due to the presence of the approximation error. Once the solution is not optimal with respect to the selected cost function, it may not even stabilize the system. Therefore, stability analysis of the control resulting from the AVI is non-trivial. This subsection is aimed at this pursuit.

Let the AVI be terminated at the $i$th iteration, once a convergence tolerance, denoted with positive (semi-)definite function $\delta(x)$, is achieved, i.e., when 
\begin{equation}
 | \hat{V}^{i+1}(x) - \hat{V}^{i}(x) | \leq \delta(x), \forall x \in \Omega. \label{Conv_Criteria}
\end{equation}
Note that if approximation errors do not exist, the convergence of VI to a finite limit function, \cite{Lincol_RelaxingDynProg, AlTamimi, Heydari_TCYB}, guarantees the satisfaction of the convergence criterion (\ref{Conv_Criteria}) for a large enough, but finite $i$ for any given arbitrary positive definite $\delta(.)$,\setcounter{footnote}{0}\footnote{To be more precise, satisfaction of (\ref{Conv_Criteria}) after a finite $i$ for an arbitrary $\delta(.)$ needs \textit{uniform} convergence of exact VI. While, the cited proofs provide its pointwise convergence. However, uniform convergence also can be proved, for example assuming boundedness of $V^*(x)$ in $\Omega$, the result given in \cite{Lincol_RelaxingDynProg} leads to the desired uniform convergence, \cite{Rinehart_Dahleh_VI_Switching}.}. 
However, if the errors exist, an arbitrarily selected $\delta(.)$ can be achieved only when the approximation errors are small enough, per the uniform convergence result of Theorem \ref{Thm_UniformConv_vs_c}. 
 
Once the convergence criteria is achieved, the resulting value function $\hat{V}^{i}(.)$ can be used for calculating the feedback control, denoted with $h^{i}(.)$ through solving the minimization problem given by
\begin{equation}
		h^{i}(x) = \argmin_{u} \Big( U(x,u) + \hat{V}^{i}\big(f(x,u)\big)\Big), \label{AVI_Policy}
\end{equation}
in online operation (i.e., on the fly) based on the instantaneous state of the system, denoted with $x$. This approach, however, leads to a considerable computational load during the online operation of the system. Another approach, widely used by ADP practitioners, is training another function approximator, called \textit{actor}, for approximating the solution to the minimization problem given by Eq. (\ref{AVI_Policy}), for different states within the domain of operation. Denoting the \textit{approximation} of $h^{i}(.)$ with $\hat{h}^{i}(.)$, the approximation error of the actor, denoted with $\mu(.)$, will be introduced to the process.
\begin{equation}
		\hat{h}^{i}(x) = \argmin_{u} \Big( U(x,u) + \hat{V}^{i}\big(f(x,u)\big)\Big) + \mu(x), \forall x \in \Omega. \label{AVI_Policy_Approx}
\end{equation}
The next theorem provides a sufficient condition for asymptotic stability of $\hat{h}^{i}(.)$ in a subset of $\Omega$ which is an estimation of its region of attraction, \cite{Khalil}.
\begin{Thm} \label{Thm_Stability} 
Let the value function be approximated using a smooth function approximator with an approximation error upper bounded by $|\epsilon^{i}(x)| \leq c U(x,0), \forall x \in \Omega, \forall i \in \mathbb{N}$, for some $c \in [0,1)$. 
Also, let the Lipschitz constants of functions $U(x,.)$ and $\hat{V}^{i}\big(f(x,.)\big)$, whose existence follows from the smoothness of the functions, be given by $L_{U}$ and $L_V$. 
If the approximation error of the actor is upper bounded by 
\begin{equation}
		\|\mu(x)\|  \leq \frac{(1-c)U(x,0) -  \delta(x)}{L_{U}+L_V}, \forall x \in \Omega, \label{mu_upperbound}
\end{equation}
with the equality holding only at the origin, then the control policy $\hat{h}^{i}(.)$ resulting from the approximate value iteration, terminated with the tolerance of $\delta(x),$ 
asymptotically stabilizes the system for any initial sate selected in compact domain $\mathcal{B}_{\bar{r}} \subset \Omega$ containing the origin, where $\mathcal{B}_r:=\{ x\in\mathbb{R}^n : \hat{V}^{i}(x) \leq r \}$, $\bar{r}$ is the largest $r$ for which $\mathcal{B}_r \subset \Omega$ holds, and $\|.\|$ denotes vector norm.

\end{Thm}
\textit{Proof}: The proof is given in the appendix.

Inequality (\ref{mu_upperbound}) provides an upper bound for the norm of the actor's approximation error. However, it is important to note that the upper bound has to be positive definite, otherwise no non-zero approximation error can satisfy it. In other words, one needs the numerator of the right hand side of (\ref{mu_upperbound}) to be positive for $x \neq 0$. Therefore, it is required to have
\begin{equation}
		\delta(x) < (1-c)U(x,0), \forall x \in \Omega-\{0\}. \label{Thm2_10_2}
\end{equation}
On the other hand, one has 
\begin{equation}
 \delta(x) \leq c\big(\tilde{V}^*(x)+\tilde{\underline{V}}^*(x)\big). \label{Thm2_11}
\end{equation}
if the number of iterations of AVI is large enough, where upper bounded positive definite functions $\tilde{V}^*(x)$ and $\tilde{\underline{V}}^*(x)$ were defined in the proof of Theorem \ref{Thm_UniformConv_vs_c}. The reason is the least upper bound and the greatest lower bound of $\hat{V}^{i}(x_0)$, as $i\to\infty$, satisfy (\ref{Breve_V_4_0}) and (\ref{Breve_V_5}), given in the appendix. 
Therefore, considering inequality (\ref{Thm2_11}), if the critic's approximation error is small enough, leading to a small $c$, inequality (\ref{Thm2_10_2}) can always be achieved, which will then lead to a positive definite right hand side in (\ref{mu_upperbound}), that determines the upper bound of the actor's approximation error. 

Note that, $\delta(x)$ can be explicitly obtained from the results of the concluded AVI, e.g., $\delta(x) := | \hat{V}^{i}(x) - \hat{V}^{i+1}(x)|$. Therefore, in practice, one can check the validity of inequality (\ref{Thm2_10_2}) before training the actor and if not satisfied, will need to increase the approximation capability/richness of the critic, e.g., by increasing the number of neurons.
It is an interesting feature of the upper bound of the actor approximation error given by Theorem \ref{Thm_Stability} that it can be calculated for any general nonlinear system, because, all the parameters are either known or calculable for a given system. For example, besides checking the validity of inequality (\ref{Thm2_10_2}), which was discussed, the Lipschitz constants $L_U$ and $L_V$ can be calculated analytically or numerically through examining the trained critic and actor. Note that, in order to find the Lipschitz constants $L_U$ and $L_V$, one needs $\Gamma$, unless the functions are globally Lipschitz. Set $\Gamma$ can be obtained using the pointwise values obtained for (\ref{AVI_Policy}) and the trained actor. The former corresponds to $h^i(.)$ and the latter is $\hat{h}^i(.)$. Utilizing these data, set $\Gamma$, i.e., the union of the images of $\Omega$ under $h^i(.)$ and $\hat{h}^i(.)$, can be found (see the next section for an example).

Another interesting feature of the given stability result is the point that it admits termination of the learning process after a finite number of iterations, through admitting the convergence criterion given by (\ref{Conv_Criteria}). This holds for both cases of having or not having the approximation errors.

\section{Numerical Example: Orbital Maneuver Problem} \label{Sim}

\subsection{Problem Setup}
The orbital maneuver problem with continuous thrust simulated in \cite{Scheeres, Heydari_JGCD_Rendezvous} is selected for numerical analyses in this study. A rigid spacecraft is orbiting around the Earth. It needs to perform a maneuver to move to a given circular orbit. The regulation of the states corresponds to positioning the spacecraft in the destination orbit, with the desired velocity, to stay in the orbit after the maneuver. Assuming planar motion, the non-dimensionalized displacement vector of the center of mass of the spacecraft from the center of the orbital frame positioned at the destination orbit is denoted by $[X,Y ]^T$, where real numbers $X$ and $Y$ are the components of the vector in the orbital frame. The equations of motion of the spacecraft in the gravity field are given by \cite{Scheeres}
$$\ddot{X} - 2 \dot{Y}̇+(1+X )(1/r^3 -1) = u_X$$
$$\ddot{Y}+ 2\dot{X}+Y (1/r^3 -1)=u_Y$$ 
where $u_X$ and $u_Y$ denote the components of the non-dimensionalized total force applied on the spacecraft and $r:=\sqrt{(1+X )^2+Y^2}$. For non-dimensionalizing, a reference length, $\mathcal{R}$, and a reference time, $\mathcal{T}$, are selected. The radius of the destination orbit is selected for $\mathcal{R}$ and the inverse of the angular velocity of the spacecraft orbiting in the destination orbit, i.e., $\sqrt{(\mathcal{R}^3/\mu_E)}$, is selected for $\mathcal{T}$, where $\mu_E$ denotes the gravitational parameter for the Earth. 

Selecting the state vector as $x=[X,Y,\dot{X},\dot{Y} ]^T$ and the control vector as $u=[u_X,u_Y ]^T$, the state equation of the orbital maneuver problem can be written as
\begin{equation}
\dot{x} = 
 \left[ \begin{array}{c}  
x(3)\\
x(4)\\
2x(4)-(1+x(1) )(1/r^3-1)\\
-2x(3)-x_2 (1/r^3-1) \end{array} \right]
+  
\left[ \begin{array}{cc}  
0& 0\\
0& 0\\
1& 0\\
0& 1\end{array} \right]
u
, \label{Maneuver_Cont_Dynamics}
\end{equation}
Note that the elements of vector $x$ are denoted with $x(i)$, $i=1,2,3,4$, as opposed to the customary notation of $x_i$, to avoid mistaking them with the discrete time steps, i.e., in $x_k$ used throughout the paper.

Minimizing cost function $J = \int_{0}^{\infty} \big(100x^Tx + u^Tu\big)dt$ leads both positioning the spacecraft in the destination orbit and having it orbit with the desired orbital velocity, since both the relative position and the relative velocity will be forced to converge to zero. 

\subsection{Implementation of ADP-based Solution}
The dynamics of the problem given by (\ref{Maneuver_Cont_Dynamics}) is in the continuous-time from. Using the (non-dimensionalized) sampling time of $\Delta t = 0.01$ the continuous-time problem is discretized to 
$$x_{k+1} = F(x_k)+gu_k, \mbox{ } 
F(x):=x + \Delta t  
 \left[ \begin{array}{c}  
x(3)\\
x(4)\\
2x(4)-(1+x(1) )(1/r^3-1)\\
-2x(3)-x_2 (1/r^3-1) \end{array} \right], \mbox{ }
g := \Delta t 
\left[ \begin{array}{cc}  
0& 0\\
0& 0\\
1& 0\\
0& 1\end{array} \right].
$$
$$ Q(x) = 100 \Delta t x^Tx, R = diag(1,1,1)\Delta t.$$  

Since the system is control affine, and the utility function is quadratic in $u$, the minimum of the term in the right hand side of (\ref{AVI_1}) can be simply found by setting its gradient to zero, which leads to
\begin{equation}
		u = - \frac{1}{2} R^{-1} g^T \nabla \hat{V}^{i}\big(f(x,u)\big), \label{AVI_Policy_Simpler}
\end{equation}
where $\nabla \hat{V}^{i}(x) := \partial \hat{V}(x) / \partial x$, \cite{Heydari_GlobalOptimality}. Note that Eq. (\ref{AVI_Policy_Simpler}) is implicit, as $u$ exists on the right hand side as well. Ref. \cite{Heydari_TCYB} proves that selecting any finite $u^0$ and conducting the successive approximation given by  
\begin{equation}
		u^{j+1} = - \frac{1}{2} R^{-1} g^T \nabla \hat{V}^{i}\big(f(x,u^{j})\big), \label{AVI_Policy_Succ_Approx}
\end{equation}
$u^{j}$ converges to the solution to Eq. (\ref{AVI_Policy_Simpler}), if the sampling time, $\Delta t$, is small enough. Note that a complete set of iterations on (\ref{AVI_Policy_Succ_Approx}), called \textit{inner loop} in \cite{Heydari_TCYB}, needs to be done at each single iteration of (\ref{AVI_1}), called \textit{outer loop}. However, selecting a small enough sampling time, the inner loop iterations are observed to converge very quickly, \cite{Heydari_TCYB}. 
This approach is used for finding the minimum of (\ref{AVI_1}) during the critic training and also for training the actor using Eq. (\ref{AVI_Policy}), which due to inevitable approximation errors, leads to (\ref{AVI_Policy_Approx}).  

The linear-in-parameter structures $\hat{V}^i(.) ={W_c^i}^T \phi(.)$ and $\hat{h}^i(.) = W_a^T \sigma(.)$ are selected for function approximation, where $\phi: \mathbb{R}^n \to \mathbb{R}^{n_c}$ and $\sigma : \mathbb{R}^n \to \mathbb{R}^{n_a}$ are the nonlinear smooth basis functions to be selected, $W_c^i \in \mathbb{R}^{n_a}, \forall i$, and $W_a \in \mathbb{R}^{n_c \times m}$ are the unknown parameters to be found. Positive integers $n_c$ and $n_a$ denote the number of neurons or basis functions in the critic and the actor, respectively. It should be noted that each iteration of AVI leads to a new set of weights for the critic, i.e., the weights of the critic evolve with the iterations. Therefore, they are denoted with superscript $i$ to relate them to their respective iterations. However, only one actor will be trained to learn the resulting control policy, denoted with $\hat{h}^i(.)$. So, the actor's weight matrix, $W_a$, is not iteration dependent.

Denoting the vector whose elements are all the non-repeating polynomials made up through multiplying the polynomial elements of vector $A$ by those of vector $B$ with $A \otimes B$, the following basis functions are selected for the function approximators
\begin{equation}
	\phi(x)=[ (x\otimes x)^T,(x\otimes x\otimes x)^T ]^T, \label{selected_phi_1}
\end{equation}
\begin{equation}
	\sigma(x)=[x^T,(x\otimes x)^T ]^T. \label{selected_sigma_1}
\end{equation}

\begin{figure}[tbph]
		\centering
		\includegraphics[width=0.85\columnwidth]{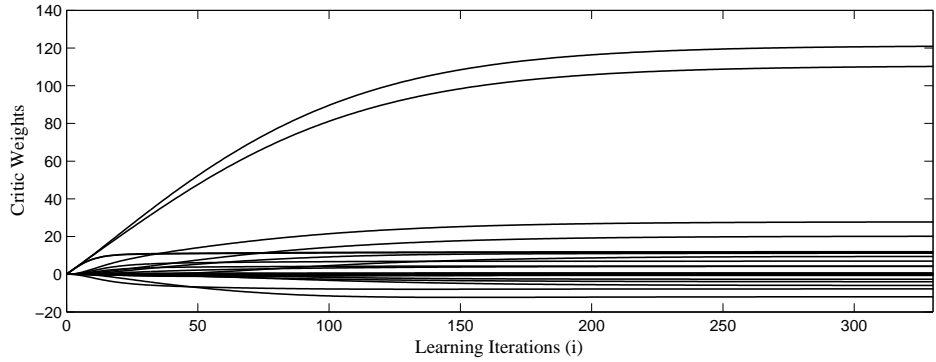} 
		\caption{Evolution of the weights of the critic during the AVI.} 
		\label{Fig_Weights}
\end{figure}

$500$ random state vectors, denoted with $x^{[p]}, p \in \{1,2,\ldots,500\}$, were selected from $\Omega_1 := \{ x\in\mathbb{R}^4:-0.3 \leq x(i)\leq 0.3, i = 1,2,3,4\}$, for learning the value function using Eq. (\ref{AVI_1}). Selecting a constant convergence tolerance of $0.01$ the convergence was evaluated in a fashion similar to (\ref{Conv_Criteria}). Starting with $\hat{V}^0(.)=0$ as the initial guess, the AVI converged after $330$ iterations of (\ref{AVI_1}), each involving an inner loop over (\ref{AVI_Policy_Succ_Approx}) which was observed to converge in less than $4$ iterations. Each iteration of the process involves finding $W_c^{i+1}$ given $W_c^i$ using
\begin{equation}
		{W_c^{i+1}}^T \phi(x^{[p]}) \approx U(x^{[p]},u^{i,[p]}) + {W_c^{i}}^T \phi \big(f(x^{[p]},u^{i,[p]})\big), \forall p \in \{1,2,\ldots,500\}, \label{AVI_1_NN}
\end{equation}
where each $u^{i,[p]}$ is the converged value of (\ref{AVI_Policy_Succ_Approx}) in which, the $x$ is substituted with the respective sample state $x^{[p]}$ and ${W_c^{i}}^T \phi(.)$ is used for $\hat{V}^{i}(.)$. The method of least squares, as detailed in \cite{Heydari_Franklin}, was used for finding $W_c^{i+1}$. 
Fig. \ref{Fig_Weights} shows the evolution of the elements of the critic's weight, versus the iteration index. 
In terms of the elapsed time, the critic training took around $80$ seconds on a desktop computer with Intel Core i7-3770, 3.40 GHz processor and 8 GB of memory, running Windows 7 and MATLAB 2013 (single threading).

Once the critic training is concluded, the actor training is done in one shot over the selected random states, that is, $W_a$ is found using
\begin{equation}
		W_a^T \phi(x^{[p]}) \approx u^{i,[p]}
		, \forall p \in \{1,2,\ldots,500\},  \label{AVI_2_NN}
\end{equation}
evaluated at $i = 330$, i.e., the iteration at which the critic training converged. 

\subsection{Analysis of the Results}

For \textit{evaluation} of the function approximation accuracy, i.e., to quantify the approximation error $\epsilon^i(x), \forall i$, \textit{another} set of sample states were selected. The point is, the approximation error at the sample states used in the training, $x^{[p]}$s, may be very low, but, it is important to evaluate the error at other states, to evaluate the \textit{generalization} accuracy of the function approximators. To this goal, $20000$ equidistant states were selected by gridding $\Omega_1$, denoted with $y^{[p]}, p \in \{1,2,\ldots,20000\}$. Function $\epsilon^i(x)$ is then given by
\begin{equation}
		\epsilon^i(y^{[p]}) := {W_c^{i+1}}^T \phi(y^{[p]}) - \Big(U(y^{[p]},w^{i,[p]}) + {W_c^{i}}^T \phi \big(f(y^{[p]},w^{i,[p]})\big)\Big), \forall p \in \{1,2,\ldots,20000\}, \forall i, \label{AVI_1_NN_eval}
\end{equation}
where $w^{i,[p]}$ is the converged value of (\ref{AVI_Policy_Succ_Approx}) evaluated at the respective sample state $y^{[p]}$.

Having the pointwise values of function $\epsilon^i(y^{[p]}), \forall p, \forall i$, 
constant $c$ used in $\|\epsilon^i(x)\| \leq cU(x,.), \forall i, \forall x$, can be found using
$$ c \approx \max_{\small {\begin{array}{c} p \in \{1,2,\ldots,20000\} \\ i \in \{1,2,\ldots,330\} \end{array}}} \frac{|\epsilon^i(y^{[p]})|}{U(y^{[p]},0)}$$
which led to $c = 0.15$. If the $c$ was obtained using $x^{[p]}$s, that is, at the states utilized in the training, the result would be $c \approx 0.10$, which is close to what was achieved using new set of states. This demonstrates the good generalization capability of the function approximator. Note that a value less than $1$ is desired per the theory presented in this work, e.g., for proof of boundedness as in Theorem \ref{Thm_Boundedness}. However, the iteration has already converged, therefore, the concern of divergence does not exist in here. But, the existing concern is the quality of the result compared with the optimal solution and the reliability of the controller.

If the assumptions of Theorem \ref{Thm_Stability} hold, asymptotic stability of the controller can be concluded. The main issue is verification of inequality (\ref{mu_upperbound}), for which, function $\mu(.)$ is needed. The process of evaluation of the approximation accuracy of the critic at the new state vectors $y^{[p]}$ are done for quantifying $\mu(.)$ as well. As for the upper bound of this error given by Eq. (\ref{mu_upperbound}), functions $U(y^{[p]},0)$ and $\delta(y^{[p]}) = |{W_c^{i}}^T \phi(y^{[p]}) - {W_c^{i-1}}^T \phi(y^{[p]})|$ evaluated at $i=330$ are used. But, the Lipschitz constants $L_U$ and $L_V$ are also required, cf. Eqs. (\ref{Thm2_3}) and (\ref{Thm2_5}) given in the appendix. Note that the respective functions are smooth, hence, differentiable. So, finding the maximum of their gradient with respect to $u$ leads to their Lipschitz constants, \cite{Marsden}. 
$$
\frac{\partial \hat{V}^i\big(f(x,u)\big)}{\partial u} = \frac{\partial \hat{V}^i(x)}{\partial x}|_{x=f(x,u)} \frac{\partial f(x,u)}{\partial u} = {W_c^i}^T \nabla \phi\big(f(x,u)\big) g 
$$
therefore
$$
L_V \approx \max_{p \in \{1,2,\ldots,20000\}} {W_c^{330}}^T \nabla \phi\big(f(y^{[p]},w^{330,[p]})\big) g = 0.186
$$
To be more accurate, $L_V$ is the maximum number between the result of the foregoing equation and 
$$
L_V \approx \max_{p \in \{1,2,\ldots,20000\}} {W_c^{330}}^T \nabla \phi\Big(f\big(y^{[p]},W_a^T\sigma(y^{[p]})\big)\Big) g
$$
where the difference is one is evaluated at $w^{300,[p]}$s and the other one at $W_a^T\sigma(y^{[p]})$s. Note that the former is $h^i(y^{[p]})$ and the latter is $\hat{h}^i(y^{[p]})$, for $i=330$. But, considering the maximum norm of the actor approximation error given by $\mu_{max} := \max_{p \in \{1,2,\ldots,20000\}} \|\mu(y^{[p]})\| = 0.02$ compared with the maximum norm of the control which was observed to be around 10, the difference between the two evaluations of $L_V$ turned out to be negligible.  

Similarly,
$
\partial U(x,u) / \partial u = 2 u^T R 
$
therefore
$$
L_U \approx \max_{p \in \{1,2,\ldots,20000\}} 2{w^{{330},[p]}}^T R \approx   \max_{p \in \{1,2,\ldots,20000\}} 2\sigma^T(y^{[p]})W_a R = 0.186
$$
Evaluating $\mu(x)$ and its upper bound given by (\ref{mu_upperbound}), it turned out that $\|\mu(x)\|$ never exceeds the bound. As a matter of fact, it remains smaller than $22\%$ of the upper bound. Therefore, the asymptotic stability of the controller about the origin follows. 

\begin{figure}[tbph]
		\centering
		\includegraphics[width=0.85\columnwidth]{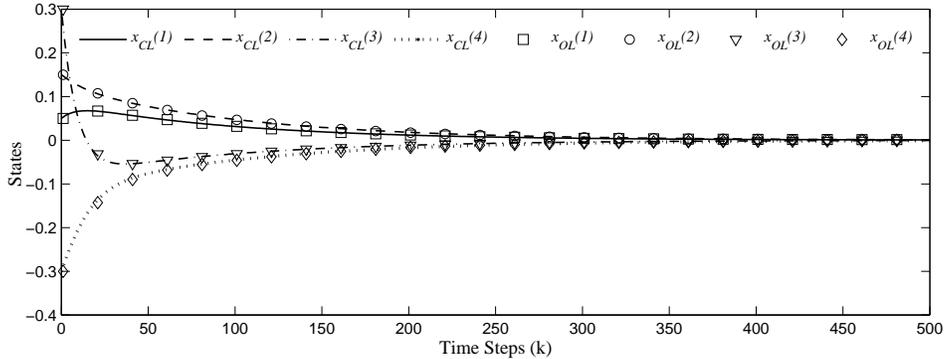} 
		\caption{Simulation results for the neurocontroller, denoted with $x_{CL}$, and for the open loop optimal controller, denoted with $x_{OL}$.} 
		\label{Fig_States}
\end{figure}

Selecting the initial condition of $x_0:=[0.05, 0.15, 0.3, -0.3]^T$, the system is operated using the trained neurocontroller and the resulting state trajectories are presented in Fig. \ref{Fig_States}. For comparison purposes, the (open loop) optimal solution to the problem is calculated numerically, using direct method of optimization, and super-imposed with the results. It can be seen from these results that the controller has been very accurate in approximating the optimal solution. 
Besides comparing the resulting state trajectories, the cost-to-go's also can be compared. The cost-to-go for the numerical open loop solution (the optimal cost-to-go) turned out to be $4.1161$, which is slightly less than the cost-to-go resulting from the close-loop controller, $4.1168$. Note that the latter is the actual resulting cost-to-go using the trained neurocontroller, not the one approximated by $\hat{V}^{330}(x_0)=4.1023$. This approximation, however, is supposed to be upper and lower bounded by the optimal cost-to-go's corresponding to cost functions (\ref{Cost_Upper_1}) and (\ref{Cost_Lower_1}), per Theorem \ref{Thm_Boundedness}. The upper and lower cost-to-go's were numerically found to be $4.6094$ and $3.6148$, respectively, which confirm the analytical result given by the theorem and provide an idea of the near optimality of the AVI results.
 
Considering the previous simulated initial conditions, it is seen that the state trajectory did not exit $\Omega$, as no state element ever exited the interval of $[-0.3, 0.3]$. Therefore, the control calculated by the neurocontroller was valid. However, this was not guaranteed or obvious from the given initial condition. But, per Theorem \ref{Thm_Stability} one can find an estimation of the ROA for the trained neurocontroller, in order to guarantee such a desired behavior.  

As for finding the estimation of the ROA, numerically analyzing $\mathcal{B}_r$ it was observed that $\bar{r} = 1.05$ for the selected $\Omega_1$, where $\bar{r}$ is the largest $r$ using which $\mathcal{B}_r \subset \Omega_1$. But, evaluating the converged critic at the selected initial condition, one has $\hat{V}^{330}(x_0) = 4.1023$, which means $x_0 \notin \mathcal{B}_{\bar{r}}.$ Therefore, it was \textit{not} guaranteed that $x_k \in \Omega_1, \forall k$. If interested to utilize the trained neurocontroller with guaranteed stability, one needs to select smaller initial conditions, such that they belong to $\mathcal{B}_{\bar{r}}$. Note that $\hat{V}^{330}(.)$ is continuous and vanishes at the origin. Therefore, $\mathcal{B}_{1.05}$ is a compact set with the origin as an \textit{interior} point, \cite{Rudin}. Details of this conclusion are given in the proof of Theorem \ref{Thm_Stability} in the appendix.
However, if controlling larger initial conditions, like the selected $x_0$, is of interest, one needs to re-train the neurocontroller using a larger domain of interest. To this purpose, $\Omega_2 := \{ x\in\mathbb{R}^4:-0.5 \leq x(i)\leq 0.5, i = 1,2,3,4\}$ was selected and the neurocontroller was retrained. Note that as the training domain is expanded it is advisable to pick more random sample states as well. For this training $2000$ random states were selected from $\Omega_2$, instead of $500$ used earlier. Once the training is concluded, evaluating the critic upper bound constant $c$ using the discussed method, it was observed to be around $0.7$, which is close to the critical value of $1$. Such a large critic approximation error led to the norm of the actor approximation error $\|\mu(x)\|$ exceeding its upper limit by $20\%$. Therefore, not only we didn't expand $\mathcal{B}_{\bar{r}}$, but also, the asymptotic stability of the origin is no longer guaranteed. Note that, simulating this controller one still gets good results as in Fig. \ref{Fig_States}, at least for that specific $x_0$, however, such a good result is not theoretically guaranteed using the presented analyses in this study, due to the violation of (\ref{mu_upperbound}).

The problem can be resolved by improving the approximation capability of the function approximators. An option is using multi-layer neural networks. Another option is using richer basis functions. For example, instead of the basis function (\ref{selected_phi_1}) and (\ref{selected_sigma_1}), one may select the richer sets of basis functions given by

 \begin{equation}
	\phi(x)=[ (x\otimes x)^T,(x\otimes x\otimes x)^T,(x\otimes x\otimes x\otimes x)^T,(x\otimes x\otimes x\otimes x\otimes x)^T ]^T, \label{selected_phi_2}
\end{equation}
\begin{equation}
	\sigma(x)=[x^T,(x\otimes x)^T,(x\otimes x\otimes x)^T,(x\otimes x\otimes x\otimes x)^T ]^T. \label{selected_sigma_2}
\end{equation}

Selecting this new set of basis functions the training was redone over $\Omega_2$ and the critic upper bound constant, $c$, turned out to be $0.26$, with $\|\mu(x)\|$ never exceeding $0.24\%$ of its upper bound. This new neurocontroller lead to $\bar{r} = 4.16$, therefore, $x_0 \in \mathcal{B}_{\bar{r}}$ and one can be assured that the trajectory will not exit the domain on which the neurocontroller is trained. Using this new neurocontroller for the given initial conditions, it was observed that the results are extremely similar to what presented in Fig. \ref{Fig_States}. This similarity may mean that the developed sufficient conditions for guaranteed stability and ROA are still conservative and milder conditions for the approximation bounds can probably be obtained. 

\section{Conclusions}
Analytical investigations of the effects of the approximation errors on the quality of the result of approximate dynamic programming were conducted. It was observed through verifiable assumptions and conditions that the learning results remain bounded. Once the learning is terminated after a finite number of iterations, it was shown that the stability of the result can be verified and an estimation of the domain of attraction can be obtained. The comprehensive numerical analysis of the theoretical results through a non-trivial fourth-order aerospace problem demonstrated the process of utilizing the theory in practice. These results lay the foundation for and push the state of the art in improving the mathematical rigor of the field of intelligent/bio-inspired control.

\appendix

The proofs of Theorems \ref{Thm_UniformConv_vs_c} and \ref{Thm_Stability} are given in this appendex.

\textit{Proof of Theorem \ref{Thm_UniformConv_vs_c}}:
Let the optimal value function associated with cost function (\ref{CostFunction}) be given by $V^*(x)$. Let $\tilde{V}^*(x)$ be defined as
\begin{equation}
\tilde{V}^*(x_0) :=\sum_{k=0}^\infty {U\big(x^{h^*}_{k}, 0\big)}, \forall x_0 \in \mathbb{R}^n, \label{Breve_V_1}
\end{equation}  
where $x_k^{h^*}:=f\big(x_{k-1}^{h^*},h^*(x_{k-1}^{h^*})\big), \forall k \in \mathbb{N}-\{0\},$ and $x_0^{h^*} := x_0$. 
In other words, the summation in (\ref{Breve_V_1}) is evaluated along the `optimal' trajectory with respect to (\ref{CostFunction}). One has
\begin{equation}
	V^*(x) \leq \overline{V}^*(x), \forall x \in \mathbb{R}^n, \label{Breve_V_2}
\end{equation}  
where $\overline{V}^*(x)$ is optimal value function associate with cost function (\ref{Cost_Upper_1}), otherwise, the control resulting from $\overline{V}^*(x)$ will be the optimal control for cost function (\ref{CostFunction}). Moreover,
\begin{equation}
	\overline{V}^*(x) \leq V^*(x) + c\tilde{V}^*(x), \forall x \in \mathbb{R}^n, \label{Breve_V_3}
\end{equation}  
otherwise $\overline{V}^*(x)$ will not be the optimal value function for cost function (\ref{Cost_Upper_1}). Note that, both sides of inequality (\ref{Breve_V_3}) include infinite sums of $U(x_k,u_k)+cU(x_k,0)$ terms, but, they are evaluated along different trajectories, i.e., the applied controls are different. The summation in the left hand side is based on the control which minimizes (\ref{Cost_Upper_1}) and the summation in the right hand side is based on the control that minimizes cost function (\ref{CostFunction}). 

By inequalities (\ref{Breve_V_2}) and (\ref{Breve_V_3}), one has
\begin{equation}
	| V^*(x) - \overline{V}^*(x)| \leq c \tilde{V}^*(x), \forall x \in \mathbb{R}^n. \label{Breve_V_4_0}
\end{equation}  
Let $\tilde{V}^*_{max}:=sup_{x\in\Omega} \tilde{V}^*(x)$. Note that $\tilde{V}^*_{max}$ is a \textit{finite} constant, by Assumption \ref{Assum_ExistingAdmissibleCont}. Therefore, the foregoing inequality leads to
\begin{equation}
	| V^*(x) - \overline{V}^*(x)| \leq c \tilde{V}^*_{max}, \forall x \in \mathbb{R}^n. \label{Breve_V_4}
\end{equation}  
Inequality (\ref{Breve_V_4}) proves the convergence of $\overline{V}^*(x)$ to the optimal value function associated with cost function (\ref{CostFunction}) as $c \to 0$. Moreover, since the right hand side of (\ref{Breve_V_4}) is independent of $x$, this convergence is uniform, \cite{Rudin}. 

Let $\tilde{\underline{V}}^*(x)$ be defined as
\begin{equation}
\tilde{\underline{V}}^*(x_0):=\sum_{k=0}^\infty {U\big(x_k^{\underline{h}^*}, 0\big)}, \forall x_0 \in \mathbb{R}^n, \label{Breve_V_4_1}
\end{equation}  
where $\underline{h}^*(.)$ is the optimal control policy for cost function $\underline{J}$,
i.e., the summation in the right hand side of (\ref{Breve_V_4_1}) is evaluated along the trajectory which is optimal with respect to $\underline{J}$ given by (\ref{Cost_Lower_1}).
Through a similar argument it can be seen that $\underline{V}^*(x) \leq V^*(x)$ and $V^*(x) \leq \underline{V}^*(x) + c \tilde{\underline{V}}^*(x)$ which leads to
\begin{equation}
	| V^*(x) - \underline{V}^*(x)| \leq c \tilde{\underline{V}}^*(x), \forall x \in \mathbb{R}^n. \label{Breve_V_5}
\end{equation}  
Defining $\tilde{\underline{V}}^*_{max}:=sup_{x\in\Omega} \tilde{\underline{V}}^*(x)$ a similar uniform convergence can be concluded as the right hand side of (\ref{Breve_V_5}) will be upper bounded by the $x$-independent term $c \tilde{\underline{V}}^*_{max}$. It should be noted that $\tilde{\underline{V}}^*_{max}$ will be finite as long as $c\in [0,1)$. The reason is the finiteness of $V^*(x), \forall x\in\Omega$ which leads to a finite $\underline{V}^*(x)$, because $\underline{V}^*(x) \leq V^*(x), \forall x\in\Omega$. One has $U(x,u)=Q(x) + u^TRu$, hence, 
\begin{equation}
\underline{V}^*(x_0)=\sum_{k=0}^\infty {\big((1-c)Q(x_k^{\underline{h}^*})+{\underline{h}^*}^T(x_k^{\underline{h}^*}) R {\underline{h}^*}(x_k^{\underline{h}^*}))\big)}
\end{equation}
being finite leads to a finite $\sum_{k=0}^\infty {(1-c)Q(x_k^{\underline{h}^*})}$ 
and the finiteness of the latter leads to a finite $\tilde{\underline{V}}^*(x_0)=\sum_{k=0}^\infty {Q(x_k^{\underline{h}^*})}$ when $0\leq c < 0$.
Finally, these uniform convergence results along with Theorem \ref{Thm_Boundedness} prove this theorem. 
$\QED$

\textit{Proof of Theorem \ref{Thm_Stability}}: The idea for the proof is using $\hat{V}^{i}(.)$ as a Lyapunov function for the system, \cite{Khalil}. 
From the boundedness of $\hat{V}^i(x)$ per Lemma \ref{Lem_Boundedness} and the positive definiteness of the bounds (they are value functions of the respective \textit{finite-horizon} cost functions as shown in \cite{Heydari_TCYB}) for $i>0$ it follows that $\hat{V}^i(x)$ is a positive definite function.

Considering (\ref{Conv_Criteria}) one has
\begin{equation}
		\hat{V}^{i}(x) + \delta(x) \geq \hat{V}^{i+1}(x) , \forall x \in \Omega, \label{Thm2_0}
\end{equation}
Using (\ref{Thm2_0}) in Eq. (\ref{AVI_1}), considering (\ref{AVI_Policy}), leads to 
\begin{equation}
		\hat{V}^{i}(x) \geq U(x,{h}^{i}(x)) + \hat{V}^{i}\big(f(x,{h}^{i}(x))\big) + \epsilon^{i}(x) - \delta(x), \forall x \in \Omega. \label{Thm2_1}
\end{equation}

Note that equation (\ref{Thm2_1}) is based on $h^i(.)$, i.e., it is independent of the actor's approximation error, see remarks at the end of section \ref{AVI}. So the next step is replacing $h^i(.)$ with $\hat{h}^i(.)$, since the system will be operated using $\hat{h}^i(.)$. 
From the Lipschitz continuity of $f(x,.), U(x,.),$ and $\hat{V}^{i}(.)$ within compact sets $\Omega$ and $\Gamma$, which follows from their smoothness in the respective compact domains \cite{Marsden}, one has
\begin{equation}
		\| U(x,u) - U(x,v) \| \leq L_{U} \| u-v \|, \forall x \in \Omega, \forall u,v \in \Gamma, \label{Thm2_3}
\end{equation}
\begin{equation}
		\| \hat{V}^{i}(f(x,u)) - \hat{V}^{i}(f(x,v)) \| \leq L_{V} \| u-v \|, \forall x \in \Omega, \forall u,v \in \Gamma, \label{Thm2_5}
\end{equation}
where $\Gamma$ is a compact subset of $\mathbb{R}^m$ such that $\hat{h}^{i}(x) \in \Gamma$ and $h^{i}(x) \in \Gamma, \forall x\in \Omega$. In other words, $\Gamma$ is the union of the images of $\Omega$ under $h^i(.)$ and $\hat{h}^i(.)$.
From inequalities (\ref{Thm2_3}) and (\ref{Thm2_5}) one has 
\begin{equation}
		U\big(x,{h}^{i}(x)\big) \geq U\big(x,{h}^{i}(x)+\mu(x)\big) - L_{U} \|\mu(x)\|, \forall x \in \Omega, \label{Thm2_6}
\end{equation}
\begin{equation}
		\hat{V}^{i}\Big(f\big(x,{h}^{i}(x)\big)\Big) \geq \hat{V}^{i}\Big(f\big(x,{h}^{i}(x)+\mu(x)\big)\Big) - L_V \|\mu(x)\|, \forall x \in \Omega. \label{Thm2_7}
\end{equation}
After replacing $U\big(x,{h}^{i}(x)\big)$ and $\hat{V}^{i}\big(f\big(x,{h}^{i}(x)\big)\big)$ in the right hand side of Eq. (\ref{Thm2_1}) using inequalities (\ref{Thm2_6}) and (\ref{Thm2_7}) one has    
\begin{equation}
		\hat{V}^{i}(x) \geq U(x,\hat{h}^{i}(x)) - L_{U} \|\mu(x)\| + \hat{V}^{i}\big(f(x,\hat{h}^{i}(x))\big) - L_V \|\mu(x)\| + \epsilon^{i}(x) - \delta(x), \forall x \in \Omega, \label{Thm2_8}
\end{equation}
because $\hat{h}^{i}(x) = {h}^{i}(x) + \mu(x)$. 

The asymptotic stability follows if 
\begin{equation}
		\Delta \hat{V}^{i}(x) := \hat{V}^{i}\big(f(x,\hat{h}^{i}(x))\big) - \hat{V}^{i}(x) \leq 0, \forall x \in \Omega, \label{Thm2_10}
\end{equation}
with the equality holding only at $x=0$. Considering (\ref{Thm2_8}), condition (\ref{Thm2_10}) holds if
\begin{equation}
		\|\mu(x)\| \leq \frac{U\big(x,\hat{h}^{i}(x)\big) + \epsilon^{i}(x) - \delta(x)}{L_{U} + L_V}, \forall x \in \Omega, \label{Thm2_9}
\end{equation}
with the possible equality only at the origin.
Using $U(x,0) \leq U(x,u), \forall u \in \Gamma$ and $|\epsilon^{i}(x)| \leq c U(x,0)$, which leads to $-cU(x,0) \leq \epsilon^{i}(x)$ one has 
\begin{equation}
		(1-c)U(x,0) \leq U\big(x,\hat{h}^{i}(x)\big) + \epsilon^{i}(x), \forall x \in \Omega. \label{Thm2_9_1}
\end{equation}
Considering (\ref{Thm2_9_1}), if inequality (\ref{mu_upperbound}) holds, then (\ref{Thm2_9}) will hold, which leads to $\Delta \hat{V}^{i}(x) \leq 0$. In the foregoing inequality the two sides are equal only at the origin, due to the positive definiteness of $U(.,.)$. Hence, value function $\hat{V}^{i}(x)$ serves as a Lyapunov function and the asymptotic stability of the system under the approximate control policy $\hat{h}^{i}(.)$, within $\Omega$, follows, as long as the entire state trajectory remains inside $\Omega$. 
Because, if it leaves $\Omega$, the control policy $\hat{h}^{i}(.)$ will no longer be valid, i.e., relation (\ref{Thm2_1}), which is the backbone of the stability result, will no longer hold. This concern can be resolved by considering the fact that $\mathcal{B}_{\bar{r}}$ will be an estimation of ROA for the system \cite{Khalil}, per the definition of $\mathcal{B}_{\bar{r}}$ and inequality $\Delta \hat{V}^{i}(x) \leq 0$ which guarantees that a state trajectory initiated within $\mathcal{B}_{\bar{r}}$ remains inside $\mathcal{B}_{\bar{r}}$. Therefore, the asymptotic stability of the control policy inside $\mathcal{B}_{\bar{r}}$ follows.

Finally, since $\mathcal{B}_{\bar{r}}$ is contained in $\Omega$, it is bounded. Also, the set is closed, because, it is the \textit{inverse image} of a closed set, namely $[0,\bar{r}]$ under a continuous function (due to the continuity of the function approximator), \cite{Rudin}. Hence, $\mathcal{B}_{\bar{r}}$ is compact. It also contains the origin, because $\hat{V}^i(0)=0$ which is the consequence of its lower and upper boundedness established in Lemma \ref{Lem_Boundedness}.
$\QED$


\begin{thebibliography}{10}

\bibitem{Watkins}
C.~Watkins, {\em Learning from Delayed Rewards}.
\newblock PhD Dissertation, Cambridge University, Cambridge, England, 1989.

\bibitem{Werbos}
P.~J. Werbos, ``Approximate dynamic programming for real-time control and
  neural modeling,'' in {\em Handbook of Intelligent Control} (D.~A. White and
  D.~A. Sofge, eds.), pp.~493--525, Multiscience Press, 1992.

\bibitem{Sutton}
R.~S. Sutton and A.~G. Barto, {\em Reinforcement Learning: An Introduction}.
\newblock MIT Press, 1998.
\newblock pp. 83-105.

\bibitem{Bertsekas_NDP}
D.~P. Bertsekas and J.~N. Tsitsiklis, {\em Neuro-Dynamic Programming}.
\newblock Athena Scientific, 1996.
\newblock ch. 2 \& 6.

\bibitem{Bala_Biega}
S.~N. Balakrishnan and V.~Biega, ``Adaptive-critic based neural networks for
  aircraft optimal control,'' {\em Journal of Guidance, Control and Dynamics},
  vol.~19, pp.~893--898, 1996.
\newblock DOI: 10.2514/3.21715.

\bibitem{Prokhorov}
D.~Prokhorov and D.~Wunsch, ``Adaptive critic designs,'' {\em IEEE Transactions
  on Neural Networks}, vol.~8, pp.~997--1007, 1997.
\newblock DOI: 10.1109/72.623201.

\bibitem{Venaya}
G.~Venayagamoorthy, R.~Harley, and D.~Wunsch, ``Comparison of heuristic dynamic
  programming and dual heuristic programming adaptive critics for neurocontrol
  of a turbogenerator,'' {\em IEEE Transactions on Neural Networks}, vol.~13,
  pp.~764--773, May 2002.
\newblock DOI: 10.1109/TNN.2002.1000146.

\bibitem{Si}
R.~Enns and J.~Si, ``Helicopter trimming and tracking control using direct
  neural dynamic programming,'' {\em IEEE Transactions on Neural Networks},
  vol.~14, no.~4, pp.~929--939, 2003.
\newblock DOI: 10.1109/TNN.2003.813839.

\bibitem{AlTamimi}
A.~Al-Tamimi, F.~Lewis, and M.~Abu-Khalaf, ``Discrete-time nonlinear {HJB}
  solution using approximate dynamic programming: Convergence proof,'' {\em
  IEEE Transactions on Systems, Man, and Cybernetics, Part B: Cybernetics},
  vol.~38, pp.~943--949, Aug 2008.
\newblock DOI: 10.1109/TSMCB.2008.926614.

\bibitem{Zhao_Xu_Jagannathan}
Q.~Zhao, H.~Xu, and S.~Jagannathan, ``Optimal control of uncertain quantized
  linear discrete-time systems,'' {\em International Journal of Adaptive
  Control and Signal Processing}, 2014.
\newblock DOI: 10.1002/acs.2473.

\bibitem{Liu_TCYB}
D.~Liu and Q.~Wei, ``Finite-approximation-error-based optimal control approach
  for discrete-time nonlinear systems,'' {\em IEEE Transactions on
  Cybernetics}, vol.~43, pp.~779--789, April 2013.
\newblock DOI: 10.1109/TSMCB.2012.2216523.

\bibitem{Heydari_NN_FinHor}
A.~Heydari and S.~N. Balakrishnan, ``Fixed-final-time optimal control of
  nonlinear systems with terminal constraints,'' {\em Neural Networks},
  vol.~48, pp.~61--71, 2013.
\newblock DOI: 10.1016/j.neunet.2013.07.002.

\bibitem{Bala_Han_JGCD}
D.~Han and S.~N. Balakrishnan, ``Adaptive critic-based neural networks for
  agile missile control,'' {\em Journal of Guidance, Control, and Dynamics},
  vol.~25, pp.~404--407, 2002.
\newblock DOI: 10.2514/2.4895.

\bibitem{Ferrari_Stengel}
S.~Ferrari and R.~F. Stengel, ``Online adaptive critic flight control,'' {\em
  Journal of Guidance, Control, and Dynamics}, vol.~27, pp.~777--786, 2004.
\newblock DOI: 10.2514/1.12597.

\bibitem{JonathanHow_ADP}
B.~W. James S.~McGrew, Jonathon P.~How and N.~Roy, ``Air-combat strategy using
  approximate dynamic programming,'' {\em Journal of Guidance, Control, and
  Dynamics}, vol.~33, pp.~1641--1654, 2010.
\newblock DOI: 10.2514/1.46815.

\bibitem{Bala_Jie}
J.~Ding and S.~N. Balakrishnan, ``Intelligent constrained optimal control of
  aerospace vehicles with model uncertainties,'' {\em Journal of Guidance,
  Control, and Dynamics}, vol.~35, pp.~1582--1592, 2012.
\newblock DOI: 10.2514/1.54505.

\bibitem{Heydari_JGCD_Rendezvous}
A.~Heydari and S.~Balakrishnan, ``Adaptive critic based solution to an orbital
  rendezvous problem,'' {\em Journal of Guidance, Control, and Dynamics},
  vol.~37, pp.~344--350, 2014.
\newblock DOI: 10.2514/1.60553.

\bibitem{LewisContSystMag}
F.~Lewis, D.~Vrabie, and K.~Vamvoudakis, ``Reinforcement learning and feedback
  control: Using natural decision methods to design optimal adaptive
  controllers,'' {\em IEEE Control Systems Magazine}, vol.~32, pp.~76--105, Dec
  2012.
\newblock DOI: 10.1109/MCS.2012.2214134.

\bibitem{Landelius_PhDThesis}
T.~Landelius, {\em Reinforcement learning and distributed local model
  synthesis}.
\newblock {PhD} Dissertation, Linkoping Univ., Linkoping, Sweden, 1997.

\bibitem{Bala_Liu_Convergence}
X.~Liu and S.~Balakrishnan, ``Convergence analysis of adaptive critic based
  optimal control,'' in {\em Proceedings of the American Control Conference},
  vol.~3, pp.~1929--1933, 2000.
\newblock DOI: 10.1109/ACC.2000.879538.

\bibitem{Lincol_RelaxingDynProg}
B.~Lincoln and A.~Rantzer, ``Relaxing dynamic programming,'' {\em IEEE
  Transactions on Automatic Control}, vol.~51, pp.~1249--1260, Aug 2006.
\newblock DOI: 10.1109/TAC.2006.878720.

\bibitem{Heydari_TNN_FinHor}
A.~Heydari and S.~N. Balakrishnan, ``Finite-horizon control-constrained
  nonlinear optimal control using single network adaptive critics,'' {\em IEEE
  Trans. Neural Netw. Learning Syst.}, vol.~24, no.~1, pp.~145--157, 2013.
\newblock DOI: 10.1109/TNNLS.2012.2227339.

\bibitem{Heydari_TCYB}
A.~Heydari, ``Revisiting approximate dynamic programming and its convergence,''
  {\em IEEE Transactions on Cybernetics}, vol.~44, pp.~2733-2743, 2014.

\bibitem{Singh_Discounted_ADP_ErrorAnalysis}
S.~P. Singh and R.~C. Yee, ``An upper bound on the loss from approximate
  optimal-value functions,'' {\em Mach. Learn.}, vol.~16, no.~3, pp.~227--233,
  1994.

\bibitem{Szepesv_AVI_API_ErrorAnalysis}
A.~Farahmand, C.~Szepesv\'{a}ri, and R.~Munos, ``Error propagation for
  approximate policy and value iteration,'' in {\em Advances in Neural
  Information Processing Systems} (J.~Lafferty, C.~Williams, J.~Shawe-Taylor,
  R.~Zemel, and A.~Culotta, eds.), pp.~568--576, 2010.

\bibitem{Szepesv_AVI_ErrorAnalysis}
R.~Munos and C.~Szepesv\'{a}ri, ``Finite-time bounds for fitted value
  iteration,'' {\em J. Mach. Learn. Res.}, vol.~9, pp.~815--857, 2008.

\bibitem{Vamvoudakis_Automatica}
K.~G. Vamvoudakis and F.~L. Lewis, ``Online actor-critic algorithm to solve the
  continuous-time infinite horizon optimal control problem,'' {\em Automatica},
  vol.~46, no.~5, pp.~878 -- 888, 2010.
\newblock DOI: 10.1016/j.automatica.2010.02.018.

\bibitem{Dierks_Jagannathan_Approx_Error}
T.~Dierks and S.~Jagannathan, ``Online optimal control of nonlinear
  discrete-time systems using approximate dynamic programming,'' {\em Journal
  of Control Theory and Applications}, vol.~9, no.~3, pp.~361--369, 2011.
\newblock DOI: 10.1007/s11768-011-0178-0.

\bibitem{Vamvoudakis_Vrabie_Lewis}
K.~G. Vamvoudakis, D.~Vrabie, and F.~L. Lewis, ``Online adaptive algorithm for
  optimal control with integral reinforcement learning,'' {\em International
  Journal of Robust and Nonlinear Control}, vol.~24, no.~17, pp.~2686--2710,
  2014.

\bibitem{Heydari_SAVI}
A.~Heydari, ``Stabilizing value iteration with and without approximation
  errors,'' available at arXiv:1412.5675.

\bibitem{Khalil}
H.~Khalil, {\em Nonlinear Systems}.
\newblock Prentice-Hall, 2002.
\newblock pp. 111-181.

\bibitem{Kirk}
D.~E. Kirk, {\em Optimal control theory; an introduction}.
\newblock Prentice-Hall, 1970.

\bibitem{Werbos2012}
P.~J. Werbos, ``Reinforcement learning and approximate dynamic programming
  {(RLADP)}-foundations, common misconceptions, and the challenges ahead,'' in
  {\em Reinforcement Learning and Approximate Dynamic Programming for Feedback
  Control} (F.~L. Lewis and D.~Liu, eds.), pp.~1--30, John Wiley \& Sons, 2012.

\bibitem{Weierstrass_Theorem}
H.~Jeffreys and B.~S. Jeffreys, ``Weierstrass's theorem on approximation by
  polynomials,'' in {\em Methods of Mathematical Physics}, pp.~446--448,
  Cambridge University Press, 3rd~ed., 1988.

\bibitem{Hornik_NN_Continuity}
K.~Hornik, M.~Stinchcombe, and H.~White, ``Multilayer feedforward networks are
  universal approximators,'' {\em Neural Networks}, vol.~2, no.~5,
  pp.~359--366, 1989.
\newblock DOI: 10.1016/0893-6080(89)90020-8.

\bibitem{Heydari_Franklin}
A.~Heydari and S.~Balakrishnan, ``Optimal switching between autonomous
  subsystems,'' {\em Journal of the Franklin Institute}, vol.~351, 2014.
\newblock DOI: 10.1016/j.jfranklin.2013.12.008.

\bibitem{Rudin}
W.~Rudin, {\em Principles of Mathematical Analysis}.
\newblock McGraw-Hill, 3rd~ed., 1976.
\newblock pp. 55, 87, 147.

\bibitem{Rinehart_Dahleh_VI_Switching}
M.~Rinehart, M.~Dahleh, and I.~Kolmanovsky, ``Value iteration for (switched)
  homogeneous systems,'' {\em IEEE Transactions on Automatic Control}, vol.~54,
  no.~6, pp.~1290--1294, 2009.

\bibitem{Scheeres}
C.~Park, G.~Guibout, and D.~Scheeres, ``Solving optimal continuous thrust rendezvous
  problems with generating functions,'' {\em Journal of Guidance, Control, and
  Dynamics}, vol.~29, no.~2, pp.~321--331, 2006.
\newblock DOI: 10.2514/1.14580.

\bibitem{Heydari_GlobalOptimality}
A.~Heydari and S.~N. Balakrishnan, ``Global optimality of approximate dynamic
  programming and its use in non-convex function minimization,'' {\em Applied
  Soft Computing}, vol.~24, pp.~291--303, 2014.
\newblock DOI: 10.1016/j.asoc.2014.07.003.

\bibitem{Marsden}
J.~E.~Marsden, T.~Ratiu, and R.~Abraham, {\em Manifolds, Tensor Analysis, and
  Applications}.
\newblock Springer-Verlag, 3rd~ed., 2001.
\newblock pp. 27, 74.

\end{thebibliography}
\end{document}